\documentclass[twocolumn,aps,prl,superscriptaddress,showpacs,
secnumarabic
]{revtex4}
\usepackage{bm,amsmath,amsfonts,amssymb,
esint,graphicx,color}

\newcommand{\Eq}[1]{Eq.~\eqref{#1}}
\newcommand{\eq}[1]{\eqref{#1}}
\newcommand{\Fig}[1]{Fig.~\ref{#1}}

\newcommand{\beq}{\begin{equation}}
\newcommand{\eeq}{\end{equation}}
\newcommand{\beqa}{\begin{eqnarray}}
\newcommand{\eeqa}{\end{eqnarray}}
\newcommand{\Beqa}{\begin{eqnarray*}}
\newcommand{\Eeqa}{\end{eqnarray*}}
\newcommand{\nn}{\nonumber}

\newcommand{\past}{{\phantom{\ast}}}
\newcommand{\sgn}{\text{sgn}}
\def\Xint#1{\mathchoice
   {\XXint\displaystyle\textstyle{#1}}%
   {\XXint\textstyle\scriptstyle{#1}}%
   {\XXint\scriptstyle\scriptscriptstyle{#1}}%
   {\XXint\scriptscriptstyle\scriptscriptstyle{#1}}%
   \!\int}
\def\XXint#1#2#3{{\setbox0=\hbox{$#1{#2#3}{\int}$}
     \vcenter{\hbox{$#2#3$}}\kern-.5\wd0}}
\newcommand{\pvint}{\Xint-}
\begin{document}

\title{Solitons in a one-dimensional Wigner crystal}

\author{M. Pustilnik}
\affiliation{School of Physics, Georgia Institute of Technology, Atlanta, Georgia 30332, USA}
\author{K. A. Matveev}
\affiliation{Materials Science Division, Argonne National Laboratory, Argonne, Illinois 60439, USA}

\begin{abstract}
In one-dimensional quantum systems with strong long-range repulsion particles arrange in a quasiperiodic chain, the Wigner crystal. We demonstrate that besides the familiar phonons, such one-dimensional Wigner crystal supports an additional mode of elementary excitations, which can be identified with solitons in the classical limit. We compute the corresponding excitation spectrum and argue that the solitons have a parametrically small decay rate at low energies. We discuss implications of our results for the behavior of the dynamic structure factor. 
\end{abstract}

%\date{September 21, 2014}


\maketitle
%%%%%%%%%%%%%%%%%%%%%%%%%%%%%%%%%%%%%%%

Landau's concept of elementary excitations plays a central role in our understanding of interacting quantum systems \cite{Nozieres}. Even if the interaction between the constituent particles in the system of interest is strong, low-energy excited states can be described in terms of weakly interacting elementary excitations. In this paper we study elementary excitations of a one-dimensional quantum system with strong long-range repulsion. Properties of such systems are dominated by the interaction, and can often be understood from semiclassical considerations. For example, the particles, regardless their statistics, are expected to form a configuration that minimizes the potential energy. Such minimal-energy configuration is, obviously, an equidistant chain~\cite{Meyer,Schulz,MAP,LMP}.  In the case of electrons interacting via the Coulomb potential such periodic structures are usually referred to as Wigner crystals~\cite{Wigner,Meyer}; here we adopt this term for systems of particles of any nature with strong long-range repulsion. Although quantum fluctuations destroy the long-range order in the one-dimensional Wigner crystal~\cite{Schulz}, the distances between neighboring particles remain close to their mean value $1/n_0$, where $n_0$ is the particle density. 

Classical one-dimensional Wigner crystals support propagation of harmonic waves of density. Their dispersion relation at low wave numbers $q\ll n_0$ reads
\beq
\omega(q) = vq\bigl[1 - \chi (q/n_0)^2 \bigr],
\label{1}
\eeq
where $v$ is the sound velocity and $\chi$ is a positive dimensionless coefficient that depends on the functional form of the interaction potential, but not on its strength~\cite{LMP}. In a quantum system, the wave with frequency $\omega$ and wave number $q$ corresponds to a phonon with energy $\varepsilon_\text{ph} = \hbar\omega$ and momentum $p = \hbar q$. 
The phonon spectrum \mbox{$\varepsilon_\text{ph}(p) = \hbar\omega(p/\hbar)$} is a concave function of $p$. Therefore, energy and momentum conservation laws forbid interaction-induced decay of phonons at zero temperature~\cite{LMP,MAP}. 

The nonlinear correction in the phonon spectrum $\varepsilon_\text{ph}(p)$ is small, and can often be neglected, which amounts~\cite{Schulz} to the Luttinger liquid~\cite{Haldane_LL} approximation. It is well known~\cite{ISG_RMP} that the interaction between phonons in the Luttinger liquid, although irrelevant in the renormalization group sense~\cite{Haldane_LL}, leads to divergences in perturbation theory~\cite{ISG_RMP}.  This difficulty is resolved~\cite{Rozhkov} by describing the system in terms of effective spinless fermions rather than phonons. Accordingly, elementary excitations at $p\to 0$ are fermionic quasiparticles and quasiholes~\cite{Rozhkov} with energies given by
\beq
\varepsilon_\pm(p) = vp \pm \frac{p^2\,}{2m_\ast}.
\label{2}
\eeq 
Here $m_\ast$ is the effective mass \cite{Rozhkov,Pereira,ISG_RMP}, which can be estimated as $m_\ast \sim m\sqrt{K}$~\cite{LMP}, where $m$ is the mass of the constituent particles and $K = \pi\hbar n_0/mv$ is a dimensionless parameter characterizing the interaction strength. For the Wigner crystal $K\ll 1$.

Similar to phonons, the spectrum of the quasiholes $\varepsilon_-(p)$ is a concave function of $p$, hence the quasiholes do not decay at zero temperature. It is therefore natural to view the phonons and the quasiholes as the same branch of elementary excitations, but in different regimes. The crossover between these regimes occurs at momenta of order $p_\ast$ defined by the equation \mbox{$\varepsilon_\text{ph}(p_\ast) = \varepsilon_-(p_\ast)$}, which yields the estimate $p_\ast\sim\hbar n_0\sqrt{K}$~\cite{LMP}. The crossover separates the classical regime at $p\gg p_\ast$ from the quantum regime at $p\ll p_\ast$. Indeed, unlike phonons, the fermions do not allow for a classical interpretation. Note also that the wave number corresponding to the crossover momentum, $p_\ast/\hbar$, vanishes in the classical limit $K\propto\hbar\to 0$, leaving no room for the quantum regime. 

A new element in the quantum regime $p\ll p_\ast$ is the emergence of the second excitation branch, the quasiparticle excitation with spectrum $\varepsilon_+(p)$, see \Eq{2}.  It is then natural to ask whether the Wigner crystal supports a second, distinct from the phonons, excitation mode at relatively high momenta $p\gtrsim p_\ast$, beyond the range of applicability of \Eq{2}. The main goal of this paper is to show that such excitations indeed exist and can be interpreted as solitons on the classical side of the quantum-to-classical crossover.  

We model our strongly interacting one-dimensional quantum system by the Hamiltonian 
\beq
H = \sum_l \frac{\,p^2_l}{2m\,} + \frac{1}{2}\sum_{\,l \,\neq\, l'} V(x_l - x_{l'}).
\label{3}
\eeq
Here $x_l$ and $p_l$ are the coordinate and momentum of the $l$th particle satisfying the usual commutation relations $\bigl[x_l,p_{l'}\bigr] = i\hbar\,\delta_{l,l'}$. We assume periodic boundary conditions and consider the thermodynamic limit when both the number of particles $N_0$ and the system size $L_0$ are taken to infinity, with the density $n_0 = N_0/L_0$ kept fixed. 

For excitations with wavelengths much larger than the distance between the particles $1/n_0$, including excitations with momenta of order $p_\ast\ll\hbar n_0$, the Wigner crystal can be treated as a continuous medium. Such continuum description is obtained by expanding the Hamiltonian \eq{3} in powers of the displacements $u_l = x_l - l/n_0$ and replacing the sums over $l$ and $l'$ by integrals. Substituting \mbox{$u_{l'} - u_{l} = (l' - l)\partial_l u + \frac{1}{2}(l' - l)^2\partial_l^2 u + \ldots$}, one obtains the gradient expansion $H = H_0 + H_1 +\ldots$. The leading term in this expansion corresponds~\cite{Schulz} to the Luttinger liquid~\cite{Haldane_LL} approximation. Changing the integration variable to $y= l/n_0$, we write this term as
\beq
H_0 = \int\!dy\left[\frac{\,p^2}{2mn_0} + \frac{\,mn_0v^2}{2\,}(\partial_y u)^2\right],
\label{4}
\eeq 
where the displacement field $u$ and the conjugate momentum density $p$ satisfy $\bigl[u(y),p(y')\bigr] = i\hbar\,\delta(y - y')$.

The Hamiltonian \eq{4} describes the strongly interacting quantum fluid in terms of the Lagrangian variables \cite{Landau,MA}, in which the position of the fluid element is specified by the reference coordinate $y$ rather than by the physical coordinate $x(y) = y + u(y)$. A subtle point in this description is the form of the momentum operator. The total momentum $\mathcal P = \int\!dy\,p(y)$ can be written as a sum of two terms, $\mathcal P = P + P_0$. Here 
\beq
P = -\int\!dy\,(\partial_y u)\,p(y)
\label{5}
\eeq
is the continuum version of the quasimomentum~\cite{quasi}, and $P_0$ accounts for the reciprocal lattice vector of the one-dimensional Wigner crystal; its eigenvalues are integer multiples of $2\pi \hbar n_0$. In the continuum description excitations with wavelengths of order $1/n_0$, responsible for the umklapp scattering~\cite{MAP}, are neglected, and both $P$ and $P_0$ commute with the low-energy Hamiltonian. Excitations near zero momentum ground state correspond to  $P_0 = 0$, which gives $\mathcal P = P$ for the total momentum. 

It is convenient to write $u$ and $p$ as
\beq
u = -\,\frac{\sqrt{K\,}}{\,2\pi n_0}(\varphi_+ + \varphi_-),
\quad
p = \frac{\,\hbar n_0}{2\sqrt{K\,}}\,\partial_y(\varphi_+ - \varphi_-),
\label{6}
\eeq
where the right/left-moving bosonic fields $\varphi_\pm$ satisfy $[\varphi_+,\varphi_-] = 0$ and  $[\varphi_\pm(y),\varphi_\pm(y')] = \pm\,i\pi\,\sgn(y-y')$. Substitution into Eqs.~\eq{4} and \eq{5} yields
\beq
H_0 = v(P_+ - P_-), 
\quad
P = P_+ + P_-,
\label{7}
\eeq
where $P_\pm = \pm\,\frac{\hbar}{4\pi}\!\int\!dy(\partial_y\varphi_\pm)^2$ are the momenta of the right/left-moving excitations. 

Nonlinear corrections to spectra in Eqs.~\eq{1} and \eq{2} arise due to higher-order terms in the gradient expansion. The two leading contributions of this type read 
\beq
H_1 = \frac{\hbar^2}{12\pi m_\ast}\!\int\!dy
\Bigl[(\partial_y\varphi)^3 - a_\ast(\partial^2_y\varphi)^2\Bigr],
\label{8}
\eeq
where $\varphi = \varphi_+ +\varphi_-$, $m_\ast$ is the effective mass \cite{Rozhkov}, and $a_\ast$ the emergent length scale. For the interaction potential $V(x)$ in \Eq{3} decaying as $1/x^3$ or faster the length scale $a_\ast$ is finite and can be estimated as $a_\ast \sim (n_0\sqrt{K\,})^{-1}$ ~\cite{supplemental}. The two terms in the right-hand side of \Eq{8} describe the leading nonlinearity and dispersion, respectively. The first term in \Eq{8} has lower scaling dimension and thus represents the leading irrelevant correction to $H_0$ at small momenta $p\ll \hbar/a_\ast$. Moreover, in order to obtain the leading nonlinear corrections to the excitation spectra, it is sufficient~\cite{Rozhkov} to retain in $H_1$ contributions proportional to $(\partial_y \varphi_\pm)^3$. With this approximation, $H_0 + H_1$ can be rewritten~\cite{Rozhkov,Mattis} in terms of effective non-interacting fermions, which leads to \Eq{2} for the spectra of the elementary excitations. Conversely, at relatively large momenta $p\gg \hbar/a_\ast$ it is the dispersion that has the dominant effect. With the nonlinearity term in \Eq{8} neglected, the Hamiltonian $H_0 + H_1$  is quadratic, and one finds \Eq{1} with $\chi = \frac{1}{3\pi}K(m/m_\ast)(a_\ast n_0)$, resulting in $p_\ast = 3\hbar/2a_\ast$ for the crossover momentum.

To study the crossover between the quantum and classical regimes, we focus on momenta of order $p_\ast$, where the nonlinearity and dispersion contributions to \Eq{8} have a comparable effect. With this in mind, we change the integration variable to $\xi = y/a_\ast$, and write the gradient expansion of the low-energy Hamiltonian as~\cite{supplemental}
\beq
H = vp_\ast\bigl( h_0 + \zeta h_1 + \zeta^2 h_2 +\ldots\bigr), 
\quad
\zeta = \frac{\,p_\ast}{2m_\ast v\,}.
%\quad
%\zeta = \frac{\varepsilon_\ast}{v p_\ast},
\label{9}
\eeq
Here $h_0$ and $h_1$ follow directly from Eqs.~\eq{4} and \eq{8}, respectively, and have a universal, i.e., model-independent, form. The operators $h_0$ and $h_1$ are given by integrals of \mbox{$(\partial_\xi\varphi_+)^2 + (\partial_\xi\varphi_-)^2$} and \mbox{$(\partial_\xi\varphi)^3 - (\partial^2_\xi\varphi)^2$}, respectively.
On the other hand, the operator $h_2$ consists of integrals of $(\partial_\xi\varphi)^4$, $(\partial_\xi\varphi)^2(\partial^3_\xi\varphi)$, and $(\partial^3_\xi\varphi)^2$ with model-dependent coefficients of order unity~\cite{supplemental}. 
The parameter $\zeta$ in \Eq{9} characterizes the relative magnitude of the nonlinear corrections to the excitation spectra at the quantum-to-classical crossover. At small $K$ the ratio $\zeta/K$ depends on the functional form of the interaction potential in \Eq{3}, but is independent of its strength~\cite{supplemental}. Careful analysis~\cite{supplemental} shows that the expansion \eq{9} is justified provided that both $K$ and $\zeta$ are small.
 
Consider now a state with a single right-moving excitation, such that $\langle P_+\rangle\sim p_\ast$ and $\langle P_-\rangle = 0$. In this state the expectation values of the operators $h_n$ are of order unity for all $n$. Equation \eq{9} then yields the expansion of the energy in powers of $\zeta$. Keeping the first two terms in this expansion is sufficient to lift the degeneracy between the two excitation branches. These terms correspond to the model-independent contributions $h_0$ and $h_1$ in the expansion \eq{9}. Therefore, with corrections of order $vp_\ast\zeta^2$ neglected, the excitation spectra can be written as
\beq
\varepsilon_\pm (p) = vp + \frac{p^2_\ast\,}{2m_\ast}\,e_\pm(p/p_\ast).
\label{10}
\eeq
The crossover functions $e_\pm(s)$ in \Eq{10} are the same for all models that admit the expansion \eq{9}. This universality in the main result of our paper.

Because of their universality, it is sufficient to compute the functions $e_\pm(s)$ for any model that has a Wigner crystal limit. Here we consider the hyperbolic Calogero-Sutherland model~\cite{Calogero,Sutherland} 
\beq
V(x) = \frac{\,\hbar^2}{m a_0^2}
 \frac{\lambda(\lambda - 1)}{\sinh^2\bigl(x/a_0\bigr)}
\label{11} 
\eeq
in the regime $\lambda \gg e^{\alpha}$, where $\alpha = (a_0 n_0)^{-1} \gg 1$. In this regime both $K = \pi e^\alpha(4\alpha^2\lambda)^{-1}$ and $\zeta = 3e^\alpha (8\pi\lambda)^{-1}$~\cite{supplemental} are small, which guarantees the applicability of the expansion \eq{9}. 
The model is integrable~\cite{Calogero,Sutherland}, and its excitation spectra can be found exactly by asymptotic Bethe ansatz~\cite{Sutherland}. Evaluation of the spectra at $p\sim p_\ast$ proceeds in the same fashion as a similar calculation for the Lieb-Liniger model~\cite{PM_bosons} and results in
the crossover functions $e_\pm(s)$ in parametric form,
\beq
s(\tau) = \pm\!\int_0^{\pm\,\tau}\!\!\!dt\,f(t),
\quad
e_\pm(\tau) = \frac{2\pi}{3}\!\int_0^{\pm\,\tau}\!\!\!dt\,s(t),
\label{12}
\eeq 
where $\tau > 0$. The function $f(t)$ in \Eq{12} is analytic at all real $t$ and is given by
\beq
f(t) = \frac{1}{3\sqrt{2\pi}}\!
\int_0^\infty\!\!\frac{dz\,}{z^{1/2}}
\sin(2\pi z)\Gamma(z)\,e^{-z(\ln z -1- 2\pi t)} 
\label{13}
\eeq
at $t < 0$ and 
\beq
f(t) = 
\frac{1}{3\sqrt{2\pi}}\,\,\pvint_0^\infty\!\!\frac{dz\,}{z^{3/2}}\!
\left[1 - \frac{\pi e^{z(\ln z -1 - 2\pi t)}}{\tan(\pi z)\Gamma(z)}\right]
\label{14}
\eeq
at $t > 0$. Simple poles in the integrand of \Eq{14} are understood as Cauchy principal values. On the quantum side of the crossover Eqs.~\eq{12}-\eq{14} yield
\beq
e_\pm(s) = \pm\,s^2 - \frac{1}{3}s^3 + \ldots,
\quad
s\ll 1,
\label{15}
\eeq
in agreement with \Eq{2}. On the classical side of the crossover we find
\begin{subequations}
\beqa
e_+(s) &=& \frac{3}{5}\!\left(\frac{2\pi}{3}\right)^{2/3}\!
s^{5/3} - \frac{2}{9}s  + \ldots,
\quad
s\gg 1,
\qquad
\label{16a} \\
e_-(s) &=& - \,s^3 - \frac{2}{3}s  + \ldots,
\quad
s\gg 1.
\label{16b}
\eeqa
\end{subequations}

The first terms in the right-hand sides of Eqs.~\eq{16a} and \eq{16b} have a purely classical origin, whereas the second ones represent the leading quantum corrections. The classical contributions can be obtained~\cite{Sutherland} by solving classical equations of motion instead of resorting to Bethe ansatz. In the regime we consider, the $\sinh$ function in \Eq{11} can be approximated by exponential, and the sum in the potential energy term in \Eq{3} can be restricted to nearest neighbors. Thus, the hyperbolic Calogero-Sutherland model reduces~\cite{Sutherland} to the Toda lattice model~\cite{Toda}. The corresponding classical equation of motion, the Toda equation~\cite{Toda}, has two kinds of solutions, the harmonic waves and the solitons~\cite{Toda}. Converting solutions of the Toda equation to the excitation spectra results~\cite{Sutherland} in \Eq{10} with $e_\pm(s)$ approximated by the leading terms of the asymptotes \eq{16a} and \eq{16b}. As expected, fermionic quasiholes on the quantum side of the crossover turn to phonons on its classical side, see the discussion above. At the same time, fermionic quasiparticles morph to the classical Toda solitons. 
  
\begin{figure}[t]%[htbp]
\centering
\includegraphics[width=0.999\columnwidth]{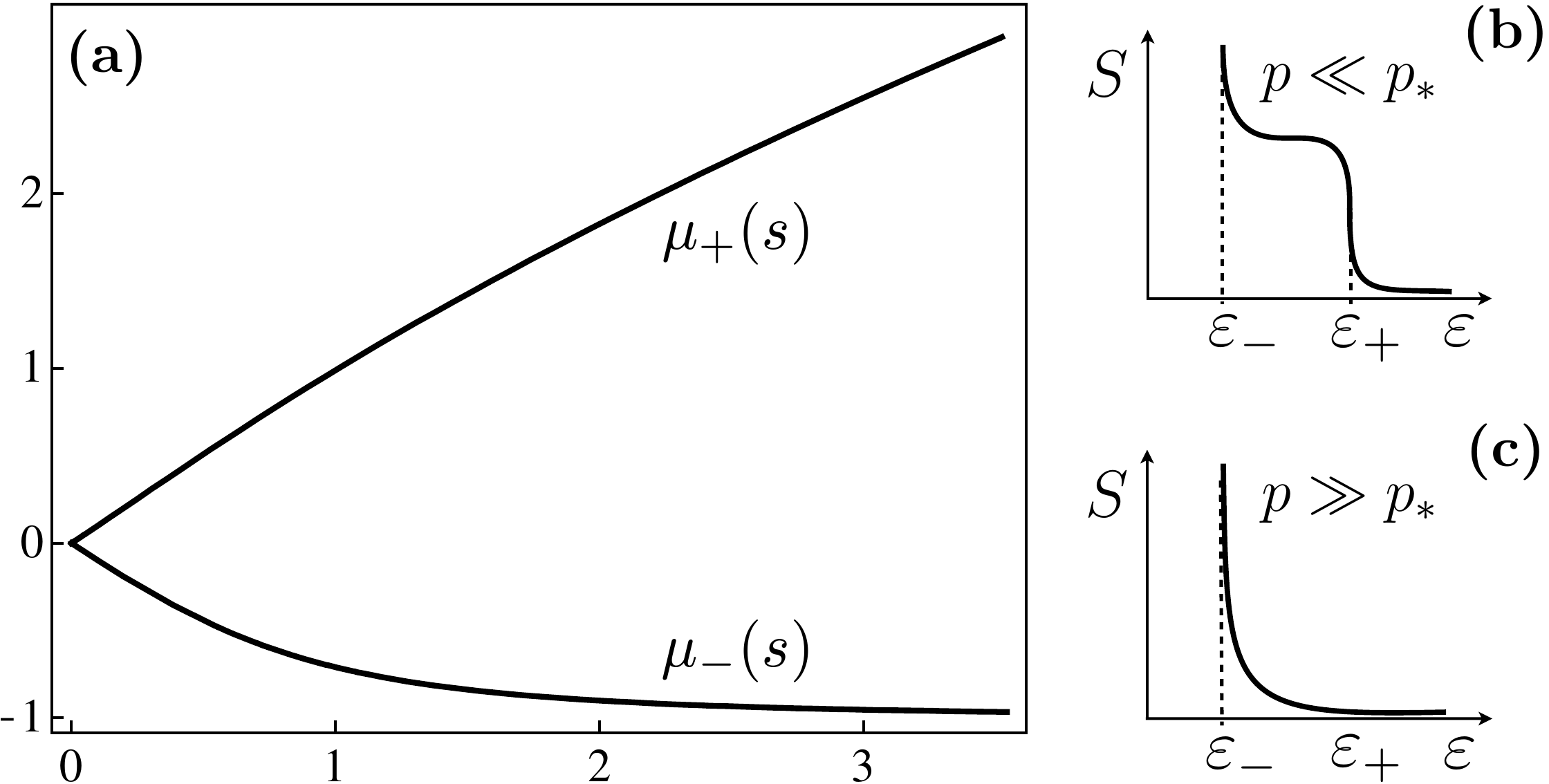}
\caption{
(a) Exponents characterizing power-law singularities in the dynamic structure factor $S(p,\varepsilon)$ at $\varepsilon\to\varepsilon_\pm(p)$, see Eqs.~\eq{17} and \eq{18}. 
(b) Sketch of the dependence of $S(p,\varepsilon)$ on $\varepsilon$ at $p\ll p_\ast$. In this regime the spectral weight is spread almost uniformly between $\varepsilon_-(p)$ and $\varepsilon_+(p)$, as expected for weakly interacting fermions.
(c) At higher $p$ the spectral weight shifts towards the phonon line $\varepsilon = \varepsilon_-(p)$, and at $p\gg p_\ast$ the dynamic structure factor $S(p,\varepsilon)$ resembles a $\delta$-function.
}
\label{Fig}
\end{figure}

The spectra $\varepsilon_\pm(p)$ reveal themselves in the behavior of the dynamic correlation functions, such as the dynamic structure factor $S(p,\varepsilon)$ defined as the Fourier transform of the density-density correlation function. At zero temperature most of the spectral weight of $S(p,\varepsilon)$ is confined between $\varepsilon_-(p)$ and $\varepsilon_+(p)$~\cite{Pereira,PKKG,ISG_RMP,Pereira09}.  Indeed, at \mbox{$\varepsilon < \varepsilon_-(p)$} the structure factor vanishes identically because $\varepsilon_-(p)$ represents the exact finite-momentum ground state of the system~\cite{PKKG,ISG_RMP}. At $\varepsilon > \varepsilon_+(p)$, on the other hand, the structure factor differs from zero due to the interaction between the right- and left-movers~\cite{drag,PKKG,Pereira09}. The corresponding coupling constant is proportional to $\zeta$, hence  $S(p,\varepsilon)$ at $\varepsilon > \varepsilon_+(p)$ is suppressed by the factor $\zeta^2$. At $\varepsilon$ approaching $\varepsilon_\pm(p)$ the structure factor exhibits power-law singularities~\cite{PKKG,ISG_RMP,Pereira09}
\beq
S(p,\varepsilon) \propto \bigl|\varepsilon - \varepsilon_\pm(p)\bigr|^{\mu_\pm(p/p_\ast)}.
\label{17}
\eeq
The exponents $\mu_\pm$ in \Eq{17} can be expressed via the spectra $\varepsilon_\pm(p)$~\cite{IG09,ISG_RMP}. Substituting $\varepsilon_\pm(p)$ in the form of \Eq{10} into  the relations derived in Refs.~\cite{IG09,ISG_RMP}, we arrive at 
\beq
\mu_\pm(s) = \left[\frac{2s}{e^\prime_\pm(s)}\right]^2 - 1,
\label{18}
\eeq
where $e^\prime_\pm(s) = de_\pm/ds$. The functions $\mu_\pm(s)$ are plotted in \Fig{Fig}(a). In the quantum regime, $s\ll 1$, \Eq{18} yields $\mu_\pm(s) = \pm\, s$, as expected for fermions with weak repulsive interaction~\cite{PKKG,ISG_RMP}. The resulting dependence of $S(p,\varepsilon)$ on $\varepsilon$ is sketched in \Fig{Fig}(b). In the classical regime, $s\gg 1$, the exponent $\mu_+$ grows as $s^{2/3}$, whereas $\mu_-$ approaches $-1$. The latter behavior is consistent with the expectation that in the classical limit the structure factor is confined to the phonon branch, \mbox{$S(p,\varepsilon)\propto\delta\bigl(\varepsilon - \varepsilon_-(p)\bigr)$}, see \Fig{Fig}(c).

%%%%%%%%%%%%%%%%%%%%%%%%%%%%%%%%
%%%%%%%%%%%%%%%%%%%%%%%%%%%%%%%%
  
In the hyperbolic Calogero-Sutherland model the quasiparticles/solitons are protected from inelastic decay by integrability: these are not merely elementary excitations, but exact eigenstates. In a generic Wigner crystal, however, these excitations acquire a finite decay rate $\Gamma/\hbar$, and the singularity at $\varepsilon\to\varepsilon_+$ in \Eq{17} is smeared by  $\Gamma$. The decay is caused by nonuniversal terms in the expansion \eq{9}, such as $h_2$. Accordingly, in the generic case the on-shell scattering amplitude for excitations with momenta $p\sim p_\ast$ is proportional to $\zeta^2$. Therefore, at $p\sim p_\ast$ the broadening $\Gamma$ is expected to be small compared with $\delta\varepsilon = \varepsilon_+ - \varepsilon_-$, which is first-order in $\zeta$. 

An estimate of $\Gamma$ can be obtained with the help of the results of Ref.~\cite{MF} which express the decay rate of the fermionic quasiparticles in terms of the corresponding spectrum. Substituting $\varepsilon_+(p)$ in the form \eq{10} with $e_+(s)$ given by \Eq{15} into the relations derived in Ref.~\cite{MF} we obtain~\cite{supplemental,KPKG} 
\beq
\Gamma(p) = g\zeta^5  vp_\ast (p/p_\ast)^8 ,
\quad
p\ll p_\ast.
\label{19}
\eeq 
The dimensionless coefficient $g$ in \Eq{19} is the functional of the interaction potential in \Eq{3}. It vanishes identically for the hyperbolic Calogero-Sutherland model~\eq{11}, but is of order unity for a generic potential~\cite{supplemental}. Extrapolating \Eq{19} to $p\sim p_\ast$ yields the estimate 
\beq
\frac{\Gamma(p_\ast)}{\delta\varepsilon(p_\ast)}\sim\zeta^4
\label{20}
\eeq
This estimate shows that the quasiparticle/soliton excitation not only remains well-defined at the quantum-to-classical crossover, but can be readily distinguished from the quasihole/phonon excitation. Moreover, it is reasonable to assume that the dependence of the ratio $\Gamma/\delta\varepsilon$ on $p$ is smooth and featureless. The estimate \eq{20} then strongly suggests that the inequality $\Gamma/\delta\varepsilon\ll 1$ holds also on the classical side of the crossover $p\gtrsim p_\ast$, breaking down at $p\sim p_{\ast\ast}\gg p_\ast$. Finding $p_{\ast\ast}$ is beyond the scope of this paper.

Our results are applicable to strongly interacting one-dimensional systems with interaction potential $V(x)$ decaying as $1/x^3$ or faster, irrespective of the statistics of the constituent particles. These results can be tested in experiments with quantum wires in the Wigner crystal regime~\cite{Meyer,exp_review}. The spectra of elementary excitations can be studied by measuring momentum-resolved tunneling between parallel quantum wires~\cite{wires_exp}, and the dynamic structure factor is accessible~\cite{drag} via measurements of the Coulomb drag effect~\cite{drag_exp}. It should be noted that in the classical regime $p\gg p_\ast$ the phonons dominate the structure factor, whereas the solitons have a negligible effect. The solitons, nevertheless, do exist in the classical regime as well. Their observation, however, would require probing the system beyond linear response. Consider, for example, the evolution in time of an initially localized density perturbation~\cite{shock_waves}. Such perturbation would break up into elementary excitations, i.e., phonons and solitons. Because the solitons propagate with supersonic velocities, they will reach remote parts of the system faster than the phonons, and their early arrival can in principle be detected in time-resolved charge transport experiments~\cite{t-resolved}. 
 
To summarize, in this paper we demonstrated that in addition to phonons, one-dimensional Wigner crystals support a second mode of elementary excitations. This mode is identified with solitons in the classical regime, and crosses over to fermionic quasiparticle excitations in the quantum regime of low momenta. The quantum-to-classical crossover in the excitation spectra is described by universal crossover functions, which we found analytically. 

\begin{acknowledgments}
This work was supported by the U.S. Department of Energy, Office of Science, Materials Sciences and Engineering Division. The authors are grateful to the Aspen Center for Physics (NSF Grant No. PHYS-1066293) for hospitality.
\end{acknowledgments}

%%%%%%%%%%%%%%%%%%%%%%%%%%%%

%%%%%%%%%%%%%%%%%%%%%%%%%%%%%

%%%%%%%%%%%%%%%%%%%%%%%%%%%%%
% this is a fake title environment
\onecolumngrid
\clearpage
%\vspace{0.8in}
\begin{center}
{
{\large \textbf{Solitons in a one-dimensional Wigner crystal}}
\\~\\
\textbf{Supplemental Material}
}

\thispagestyle{empty}

\vspace{0.15in}

M. Pustilnik$^1$ and K. A. Matveev$^2$

\vspace{0.075in}
\small\textit{
$^1$School of Physics, Georgia Institute of Technology, Atlanta, GA 30332, USA
\\
$^2$Materials Science Division, Argonne National Laboratory, Argonne, IL 60439, USA
}
\end{center}

\vspace{0.25in}

\twocolumngrid
%%%%%%%%%%%%%%%%%%%%%%%%%%%%%
\numberwithin{equation}{section}

%%%%%%%%%%%%%%%%%%%%%%%%%%%%%%%%%%%%%%%
\section*{I. Microscopic model of the Wigner crystal}
\setcounter{section}{1}
\setcounter{equation}{0}
%%%%%%%%%%%%%%%%%%%%%%%%%%%%%%%%%%%%%%%

We model our strongly interacting one-dimensional system by the Hamiltonian 
\beq
H = \sum_l \frac{\,p^2_l}{2m\,} + U,
\quad
U = \frac{1}{2}\sum_{\,l \,\neq\, l'} V(x_l - x_{l'}),
\label{1.1}
\eeq
see Eq.~(3) in the paper. For simplicity, we shall assume that the interaction potential $V(x)$ in \Eq{1.1} decreases with $|x|$ monotonically and sufficiently fast to ensure convergence of the series
\beq
V_{kk'} = \sum_{l = 1}^{\infty} V^{(k)}_l l^{k'},
\quad
V^{(k)}_l = \left.\frac{d^k V(x) }{dx^k}\right|_{x=l/n_0}.
\label{1.2}
\eeq
The quantities $V_{kk'}$ defined by \Eq{1.2} obviously satisfy
\beq
\frac{d\,}{dn_0}V_{kk'} = - \,\frac{1\,}{n_0^2}V_{k+1,k'+1}.
\label{1.3}
\eeq
We will also use the estimate
\beq
V_{k+1,k'}\sim -\,\alpha n_0 V_{kk'},
\label{1.4}
\eeq
where 
\beq
\alpha = -\,\frac{1\,}{2n_0} \left.\frac{d}{dx}\ln V(x)\right|_{x = 1/n_0}
\label{1.5}
\eeq
is a dimensionless parameter characterizing the interaction range. 

%%%%%%%%%%%%%%%%%%%%%%%%%%%%%%%%%%%%%%%
\section*{II. Low-energy Hamiltonian}
\setcounter{section}{2}
\setcounter{equation}{0}
%%%%%%%%%%%%%%%%%%%%%%%%%%%%%%%%%%%%%%%

The low-energy Hamiltonian is obtained by expanding the potential energy $U$ in \Eq{1.1}  
in powers of the displacements $u_l = x_l - l/n_0$. The leading term in the resulting expansion $U = U^{(2)} + U^{(3)} +\ldots$ is quadratic in the displacements,
\beq
U^{(2)} = %\sum_l \frac{p^2_l}{2m\,} + 
\frac{1}{4}\!\sum_{l \neq l'} V^{(2)}_{l' - l}\bigl(u_{l'} - u_l)^2.
\label{2.1}
\eeq
The harmonic approximation amounts to replacing $U$ in \Eq{1.1} with $U^{(2)}$. With this approximation, the Hamiltonian can be diagonalized exactly. It has only one branch of elementary excitations, the phonons. Their spectrum at $p\ll\hbar n_0$ has the form
\beq
\varepsilon_\text{ph}(p) = vp\bigl[1 - \chi (p/\hbar n_0)^2 + \ldots\bigr]
\label{2.2}
\eeq
with the sound velocity $v$ and the coefficient $\chi$ given by
\beq
v = \left(\frac{V_{22}\,}{m n_0^2}\right)^{\!1/2}\!,
\quad
\chi = \frac{1}{24}\frac{V_{24}}{V_{22}}.
\label{2.3}
\eeq
 
We now replace the sums over $l$ and $l'$ in \Eq{2.1} by the integrals, expand $u_{l'} - u_{l}$ in Taylor series, 
\beq
u_{l'} - u_{l} = (l' - l)\partial_l u_l + \frac{1}{2}(l' - l)^2\partial_l^2 u_l + \ldots,
\label{2.4}
\eeq 
and rescale $u_l$ as
\beq
u_l = -\,\frac{\sqrt{K\,}}{\,2\pi n_0}\varphi(y), 
\quad
y= l/n_0.
\label{2.5}
\eeq
Here $\varphi(y) = \varphi_+(y) + \varphi_-(y)$ [see Eq.~(6) in the paper], and
\beq
K = \frac{\pi\hbar n_0}{mv\,} \ll 1
\label{2.6}
\eeq
is the dimensionless parameter characterizing the interaction strength. This yields
\beq
U^{(2)} = \frac{\hbar v}{8\pi}\!\int\!dy\,(\partial_y\varphi)^2 +\ldots.
\label{2.7}
\eeq
Combining $U$ in the form of \Eq{2.7} with the kinetic energy, we obtain the Luttinger liquid Hamiltonian
\beq
H_0 = \frac{\hbar v}{4\pi}\!\int\!dy\bigl[(\partial_y\varphi_+)^2 + (\partial_y\varphi_-)^2\bigr],
\label{2.8}
\eeq
see Eq.~(7) in the paper.

Higher-order contributions responsible for the nonlinear corrections to spectra are obtained in a similar fashion. Combining the lowest order term in the gradient expansion of the cubic contribution $U^{(3)}$ with the second-order term in the gradient expansion of the quadratic contribution $U^{(2)}$, we obtain Eq.~(8) of the paper,
\beq
H_1 = \frac{\hbar^2}{12\pi m_\ast}\!\int\!dy
\Bigl[(\partial_y\varphi)^3 - a_\ast(\partial^2_y\varphi)^2\Bigr],
\label{2.9}
\eeq
where the effective mass $m_\ast$ and the emergent length scale $a_\ast$ satisfy
\beq
\frac{m_\past}{m_\ast} = -\,\frac{1}{4\sqrt{K\,}}\frac{V_{33}}{n_0V_{22}},
\quad
a_\ast n_0 = -\,\frac{\pi\,}{2\sqrt{K\,}}\frac{n_0 V_{24}}{V_{33}},
\label{2.10}
\eeq
and with the help of \Eq{1.4} can be estimated as
\beq
m_\ast \sim\frac{m\sqrt{K}}{\alpha},
\quad
a_\ast \sim \frac{1}{n_0\alpha\sqrt{K}}.
\label{2.11}
\eeq
Using \Eq{2.10}, the coefficient $\chi$ in \Eq{2.2} can be written in the form $\chi = \frac{1}{3\pi}K(m/m_\ast)(a_\ast n_0)$, quoted in the paper. Note that with the help of \Eq{1.3} the expression for the effective mass can be cast in the form
\beq
\frac{m_\past}{m_\ast} =  \frac{1}{2v \sqrt{K}} \frac{d(vn_0)}{dn_0},
\label{2.12}
\eeq
valid for all Galilean-invariant systems~[\onlinecite{mass}]. 

Changing the integration variable in Eqs.~\eq{2.8} and \eq{2.9} to $\xi = y/a_\ast$, we write the expansion of the low-energy Hamiltonian as
\beq
H = vp_\ast\bigl( h_0 + \zeta h_1 + \zeta^2 h_2 +\ldots\bigr),
\label{2.13}
\eeq
see Eq.~(9) in the paper. The expansion parameter in \Eq{2.13},
\beq
\zeta = \frac{\,p_\ast}{2m_\ast v\,},
\quad
p_\ast = \frac{3\hbar_\past}{2a_\ast},
\label{2.14}
\eeq
is given by
\beq
\zeta = \frac{3}{8\pi^2} \frac{V_{33}^2}{n_0^2 V_{22}V_{24}}K
\sim \alpha^2 K.
\label{2.15}
\eeq
The operators $h_0$ and $h_1$ in \Eq{2.13} correspond to Eqs.~\eq{2.8} and \eq{2.9}, respectively, and have a universal form
\beqa
h_0 &=& 
\frac{1}{6\pi}\int\!d\xi\Bigl[(\partial_\xi\varphi_+)^2 + (\partial_\xi\varphi_-)^2\Bigr],
\label{2.16} \\
h_1  &=& 
\frac{2}{27\pi}\!\int\!d\xi
\Bigl[(\partial_\xi\varphi)^3 - (\partial^2_\xi\varphi)^2\Bigr],
\label{2.17}
\eeqa
whereas the operator $h_2$ is given by
\beq
h_2 = \frac{4}{81\pi}\!\int\!d\xi\!\left[
a (\partial_\xi\varphi)^4 
+ b (\partial_\xi\varphi)^2(\partial^3_\xi\varphi)
+ c (\partial^3_\xi\varphi)^2
\right]\quad
\label{2.18}
\eeq
with model-dependent coefficients
\beq
a = \frac{V_{22} V_{44}}{V^2_{33}},
\quad
b = 2\,\frac{V_{22}V_{35}}{V_{24}V_{33}},
\quad
c = \frac{8}{15} \frac{V_{22} V_{26}}{V^2_{24}}.
\label{2.19}
\eeq
These coefficients are of order unity and depend on the functional form of the interaction potential, but are independent of its strength or range. Higher-order terms in the expansion \eq{2.13} have a similar structure.

\vspace{-0.1in}
%%%%%%%%%%%%%%%%%%%%%%%%%%%%%
\section*{III. Applicability of the low-energy expansion}
\setcounter{section}{3}
\setcounter{equation}{0}
%%%%%%%%%%%%%%%%%%%%%%%%%%%%%

The continuum theory is applicable for the description of excitations with momenta of order $p_\ast$ or, equivalently, with wavelengths of order $a_\ast$ provided that the length scale $a_\ast$ is large compared with the interparticle distance $1/n_0$. This leads to the condition $\alpha\sqrt{K} \ll 1$, see \Eq{2.11}, which can be also written in terms of the parameter $\zeta \sim\alpha^2 K$ as
\beq
\zeta\ll 1.
\label{3.1}
\eeq 
Alternatively, one can arrive at the condition \eq{3.1} by observing that the expansion of the potential energy $U$ in \Eq{1.1} in powers of the displacement $u_l$ is justified if $u_l$ is smaller than the scale characterizing the dependence of $U$ on $x_l$, i.e., the interaction range $1/\alpha n_0$, see \Eq{1.4},
\beq
|u_l| \ll \frac{1\,}{\alpha n_0}.
\label{3.2}
\eeq 
On the other hand, the displacements $u_l$ are bound from below by zero-point oscillations, 
\beq
|u_l| \gtrsim\frac{\sqrt{K\,}\,}{\,n_0}.
\label{3.3}
\eeq
The inequalities \eq{3.2} and \eq{3.3} are compatible only if the condition \eq{3.1} is satisfied. 

The above consideration rests on the assumption that the system can be treated as a crystal, which is possible only if the displacements are small compared with the mean interparticle distance, or, equivalently, if $K\ll 1$. For large $\alpha$ this condition is less restrictive than $\zeta\ll 1$.

\vspace{-0.1in}

%%%%%%%%%%%%%%%%%%%%%%%%%%%%%
\section*{IV. Hyperbolic Calogero-Sutherland model}
\setcounter{section}{4}
\setcounter{equation}{0}
%%%%%%%%%%%%%%%%%%%%%%%%%%%%%

Consider the potential~[\onlinecite{Sutherland_book}]
\beq
V(x) = \frac{\,\hbar^2}{m a_0^2}
\frac{\lambda(\lambda - 1)}{\sinh^2\bigl(x/a_0\bigr)}. 
\label{4.1} 
\eeq
We are interested in the dilute limit $a_0 \ll 1/n_0$, when the $\sinh$ function in \Eq{4.1} can be approximated by exponential,
\beq
V(x) \approx \frac{4\hbar^2}{m a_0^2}\,\lambda(\lambda - 1) e^{- 2|x|/a_0},
\label{4.2} 
\eeq
and \Eq{1.5} gives 
\beq
\alpha = \frac{1\,}{a_0 n_0} \gg 1.
\label{4.3}
\eeq
Because $e^{-2\alpha} \ll 1$, all but the first terms in the series \eq{1.2} can be neglected, and we find
\beq
V_{kk'} = (- \,2\alpha n_0)^k \frac{4\hbar^2\lambda(\lambda - 1)}{m a_0^2} \,e^{- 2\alpha},
\label{4.4}
\eeq
irrespective of $k'$. Using \Eq{4.4}, we obtain from Eqs.~\eq{2.3}, \eq{2.6}, and \eq{2.15}
\beq
%K = \frac{\pi e^\alpha}{4\alpha^2\sqrt{\lambda(\lambda - 1)}}\,,
%\quad
\zeta = \frac{3}{\,2\pi^2}\alpha^2 K
= \frac{3}{8\pi}\frac{\,e^\alpha}{\sqrt{\lambda(\lambda - 1)}}.
\label{4.5}
\eeq
Condition \eq{3.1} then translates to 
\beq
\lambda \gg e^\alpha.
\label{4.6}
\eeq
The inequalities \eq{4.3} and \eq{4.6} define the Toda limit~[\onlinecite{Sutherland_book}] of the hyperbolic Calogero-Sutherland model.

\vspace{0.2in}
%%%%%%%%%%%%%%%%%%%%%%%%%%%%%
\centerline{\textbf{V. Quasiparticle decay rate}}
\vspace{0.2in}

%\section*{V. Quasiparticle decay rate}
\setcounter{section}{5}
\setcounter{equation}{0}
%%%%%%%%%%%%%%%%%%%%%%%%%%%%%

The decay rate $\Gamma(p)/\hbar$ can be evaluated exactly~[\onlinecite{rate}] in the quantum regime $p\ll p_\ast$. According to Ref.~[\onlinecite{rate}],
\beq
\Gamma(p) = \frac{3}{5120\pi^3}\frac{\Lambda^2 p^8}{\hbar^4 m_\ast v^2},
\label{5.1}
\eeq
where $\Lambda$ is expressed via the quasiparticle energy  $\bar\varepsilon_+(p;n_0,\kappa)$ in a moving Wigner crystal viewed as a function of the quasiparticle momentum $p$, density $n_0$, and of the momentum of the crystal per particle $\kappa$~[\onlinecite{rate,Lagrangian}]. In terms of this energy $\Lambda$ is given by~[\onlinecite{rate}]
%\begin{widetext}
\onecolumngrid
\vspace{0.05in}
\noindent\rule{0.485\columnwidth}{0.4pt}\rule[0mm]{0.5pt}{8pt}\hfill 
\beq
\Lambda = \frac{1}{2}\left(\partial_{LR}^2\frac{1}{\bar m} - 2\pi\hbar\,\partial_L \bar\lambda\right) 
- \,\frac{\partial_L \bar v}{4\bar v}\partial_L\frac{1}{\bar m}
+ \frac{(\partial_L \bar v)^2}{4\bar m\bar v^2}
- \left( \frac{\partial_L \bar v}{4\bar v} + \frac{\bar m}{2}\partial_L\frac{1}{\bar m} \right)
\!\left(\partial_R\frac{1}{\bar m} 
-  2\pi\hbar\,\partial_L \bar\lambda\right),
\label{5.2}
\eeq
where
\beq
\bar v = \left.\frac{\,\partial \bar\varepsilon_+}{\partial p\,}\right|_{p\to 0},
\quad
\frac{1}{\bar m} = \left.\frac{\,\partial^2 \bar\varepsilon_+}{\partial p^2\,}\right|_{p\to 0},
\quad
\bar\lambda =\left.\frac{\,\partial^3 \bar\varepsilon_+}{\partial p^3\,}\right|_{p\to 0},
\label{5.3}
\eeq 
and
\beq
\partial_R = \sqrt{K}\frac{\partial\,}{\partial n_0} + \frac{\pi\hbar}{\sqrt{K}}\frac{\partial}{\partial\kappa},
\quad
\partial_L = \sqrt{K}\frac{\partial\,}{\partial n_0} - \frac{\pi\hbar}{\sqrt{K}}\frac{\partial}{\partial\kappa},
\quad
\partial_{LR}^2 = K\frac{\,\partial^2}{\partial n_0^2} - \frac{\pi^2\hbar^2}{K\,} \frac{\partial^2}{\partial\kappa^2}.
\label{5.4}
\eeq
%\end{widetext}
\hfill\rule[-7.5pt]{0.5pt}{8pt}\rule{0.475\columnwidth}{0.4pt}
\vspace{0.1in}
\twocolumngrid

For Galilean-invariant systems~[\onlinecite{Lagrangian}]
\beq
\bar\varepsilon_+(p;n_0,\kappa) = \bar\varepsilon_+(p;n_0,0) + \frac{\kappa p}{m}.
\label{5.5}
\eeq
The energy $\bar\varepsilon_+(p;n_0,0)$ coincides with $\varepsilon_+$ given by Eqs.~(10) and (15) of the paper,
\beq 
\varepsilon_+ = vp + \frac{p^2}{2m_\ast} - \frac{p^3}{6m_\ast p_\ast} + \ldots,
\quad
p\ll p_\ast.
\label{5.6}
\eeq
Eqs.~\eq{5.3}, \eq{5.5}, and \eq{5.6} then yield 
\beq
\bar v = v + \frac{\kappa}{m},
\quad
\bar m = m_\ast, 
\quad
\bar\lambda = -\,\frac{1\,}{m_\ast p_\ast}.
\label{5.7}
\eeq

The dominant contributions to $\Lambda$ in \Eq{5.2} are given by terms containing $\bar\lambda$, whereas the remaining terms are of relative order $(\zeta K)^{1/2}\ll 1$. 
Neglecting these corrections, we find
\beq
\Lambda 
%= \frac{\pi\hbar\,}{\,m_\ast} \frac{\partial\,}{\partial n_0}\frac{\sqrt{K}\,}{\,p_\ast}
= \frac{2\pi\,}{\,3m_\ast} \frac{\partial\,}{\partial n_0} \bigl(a_\ast\sqrt{K}\bigr).
\label{5.8}
\eeq
Substituting here $a_\ast$ in the form \eq{2.10} and evaluating the derivatives with the help of \Eq{1.3}, we obtain
\beq
\Lambda =  -\,\frac{\pi^2(V_{24}V_{44} - V_{33}V_{35})}{3m_\ast n_0^2 V_{33}^2}.
\label{5.9}
\eeq
Equation \eq{5.1} then yields
\beqa
\Gamma(p) = g\zeta^5v p_\ast(p/p_\ast)^8
\label{5.10}\\
\nn
\eeqa
with the coefficient $g$ given by
\beq
g = \frac{3\pi}{10}\frac{V_{22}^2} {V_{24}^2V_{33}^4}\!\left(V_{24}V_{44} - V_{33}V_{35}\right)^2.
\label{5.11}
\eeq
This coefficient depends on the functional form of the interaction potential, but not on its strength or range. As shown in the Supplemental Material to Ref.~[\onlinecite{thermalization}], \mbox{$V_{24}V_{44} = V_{33}V_{35}$} for the hyperbolic Calogero-Sutherland model \eq{4.1} irrespective of the values of $\lambda$ and $a_0$. Accordingly, in this case the coefficient $g$ vanishes identically, as expected for integrable models exhibiting no relaxation. For a generic interaction potential, however, $g$ is of order unity, and $\Gamma$ is finite.

\vspace{-0.1in}
\onecolumngrid
%%%%%%%%%%%%%%%%%%%%%%%%%%%%

%%%%%%%%%%%%%%%%%%%%%%%%%%%%

\end{document}